\documentclass[epsfig,useAMS,usegraphicx,usenatbib]{mn2e}

\def\cqr{$\chi$$^2_\nu$ }
\def\c2nor{\chi^2}
\def\eze{$E_{\rm 0}$ }
\def\epo{$E_{\rm p,obs}$ }
\def\epi{$E_{\rm p,i}$ }
\def\eiso{$E_{\rm iso}$ }
\def\ega{$E_{\gamma}$ }
\def\epeiso{$E_{\rm p,i}$ -- $E_{\rm iso}$ }
\def\epega{$E_{\rm p,i}$ -- $E_{\gamma}$ }
\def\nufnu{$\nu$F$_\nu$ }
\def\epeisotb{$E_{\rm p,i}$ -- $E_{\rm iso}$ -- $t_b$ }
\def\sax{{\it Beppo}SAX }
\def\swift{{\it Swift} }
\def\ga{\mathrel{\hbox{\rlap{\hbox{\lower4pt\hbox{$\sim$}}}\hbox{$>$}}}}
\def\la{\mathrel{\hbox{\rlap{\hbox{\lower4pt\hbox{$\sim$}}}\hbox{$<$}}}}

\title[The \epeiso correlation in GRBs: update, re--analysis and main 
implications]
{The \epeiso correlation in GRBs: updated observational status, 
re--analysis and main 
implications}
\author[L. Amati]{Lorenzo Amati$^{1}$\thanks{E-mail:
amati@iasfbo.inaf.it}\\
$^{1}$INAF - IASF Bologna, 
via P. Gobetti 101, Bologna (Italy)}
\begin{document}

\date{Submitted 2006 January 24.}

\pagerange{\pageref{firstpage}--\pageref{lastpage}} \pubyear{2005}

\maketitle

\label{firstpage}

\begin{abstract}
The correlation between the cosmological rest--frame \nufnu spectrum peak energy, \epi ,
and the
isotropic equivalent radiated energy, \eiso,
discovered by Amati et al. in 2002 and confirmed/extended by
subsequent osbervations,
is one of the
most intriguing and debated observational evidences in Gamma--Ray Bursts (GRB)
astrophysics.
In this paper I provide an update and a re--analysis of the \epeiso
correlation basing on an updated sample consisting of 41
long GRBs/XRFs with firm estimates of $z$ and observed peak energy, \epo , 12 GRBs with 
uncertain valeus of $z$ and/or \epo, 2 short GRBs with firm estimates of $z$ and \epo 
and the peculiar
sub--energetic events GRB980425/SN1998bw and GRB031203/SN2003lw. In addition to standard correlation analysis
and power--law fitting, the data analysis here reported includes
a modelization which accounts
for sample variance. All 53 classical long GRBs and XRFs, including 11
\swift events with published spectral parameters and fluences, 
have \epi and \eiso values, or upper/lower limits,
consistent with the correlation, which shows a chance probability as low as 
$\sim$7$\times$10$^{-15}$, a slope of $\sim$0.57 ($\sim$0.5 when fitting by accounting
for sample variance) and an extra--Poissonian
logarithmic dispersion of $\sim$0.15, it extends over $\sim$5 orders of magnitude 
in \eiso and $\sim$3 orders of magnitude in \epi and holds from the 
closer to the higher $z$ GRBs.
Sub--energetic GRBs (980425 and possibly 031203) and short
GRBs are found to be inconsistent with the \epeiso correlation, showing that
it can be a powerful tool for discriminating different classes of 
GRBs and understanding
their nature and differences. 
I also discuss the main implications of the updated 
\epeiso 
correlation for the models of the physics and geometry
of GRB emission, its use as a pseudo--redshift estimator and
the tests of possible selection effects with GRBs of unknown redshift.
\end{abstract}


\section{Introduction}

Since 1997, with the first discoveries of optical counterparts and host
galaxies, redshift estimates for
Gamma--Ray Bursts (GRBs) have become available,
allowing 
the study of the
intrinsic properties of this challenging astrophysical phenomena.
Among these, the correlation between the photon energy (commonly
called {\it peak energy}) at which the 
cosmological rest frame \nufnu spectrum peaks,  
\epi, and the total isotropic--equivalent radiated energy, \eiso,
is one of the most intriguing and
discussed. This correlation was
discovered by 
\cite{Amati02} based on BeppoSAX data and subsequently confirmed and
extended to X--Ray Rich GRBs (XRRs) and X--Ray Flashes (XRFs) based on HETE--2 data 
\citep{Amati03,Lamb04,Sakamoto04,Sakamoto05a}. 
It can be used to
constrain the parameters of
the various scenarios
for the physics of GRB prompt emission, it is a challenging test for jet and 
GRB/XRF (X--Ray Flashes) unification
models and it can provide
hints on the nature of different sub--classes of GRBs (sub--energetic GRBs, 
short GRBs). Also, the \epeiso correlation has been used for building up
redshift estimators and
is frequently assumed as an 
imput or as a
required output for GRB population synthesis models. 
In this paper, I provide an update (up to December 2005) 
and a re--analysis of the \epeiso correlation
based on a sample of a total of 56 events which includes  \swift GRBs
with known redhsift and published spectral parameters
and two very recent
short GRBs with firm estimates of redshift and \epi. The analysis here reported
includes also fitting the data with a model which accounts for
sample variance, given that this correlation is highly significant but also
shows a dispersion which cannot be accounted for only by statistical fluctuations
and is an important source of information.
I also discuss the various explanations and implications of the existence and 
properties of the \epeiso
correlation, its possible use for the estimate of pseudo--redshifts, and  
the tests based on GRBs with unknown
redshift aimed to the evaluation of the impact of possible selection effects. 
For this last purpouse, I also make use of published spectral parameters and
fluences of a sample of 46 HETE--2 GRBs.\\
In order to introduce some basic information for the discussion reported 
in Section 5 
and to allow a comparison between the results here reported and those reported in
previous works, the description and properties of the updated sample (Section 3)
the description and results of the
data analysis results (Section 4)
the discussion of the main implications and explanations of the \epeiso correlation 
(Section 5) and the discussion of pseudo--redshift estimates and
tests based on GRBs with unknown redshift (Section 6)
are preceeded (Section 2) by a brief review of spectral and energetics 
properties
of GRBs and of previous studies of the \epeiso correlation.

\begin{table*}
\begin{minipage}{155mm}
\caption{\epi and \eiso values for long GRBs and XRFs with firm estimates of both
$z$ and \epo (41 events),  
the peculiar sub--energetic event 
GRB980425
and the only two short GRBs with firm estimates of $z$ and \epo ,
GRB050709 and GRB051221.
The uncertainties are at 1$\sigma$ significance. 
The "Type" column
indicates wether the event is a normal long GRB (LONG), an X--Ray Flash
(XRF) or is sub--energetic (SUB--EN).
The "Instruments" column reports the name of the experiment(s), or of 
the satellite(s), that provided the estimates of spectral parameters and
fluence (BAT = BATSE, SAX = \sax, HET = HETE--2, KON = Konus, SWI = \swift). 
The last two columns report the references for the spectral
parameters and the references
for the values and uncertainties of \eiso, respectively. The "--" sign indicates
that \eiso was computed specifically for this work (see text).
GRBs detected by \swift are marked
with an asterisk.}

\begin{tabular}{llllllcc}
\hline
 GRB  & Type & $z$  & \epi  & \eiso & Instruments &  Refs. for $^{\rm (a)}$ &  
 Refs. for $^{\rm (a)}$ \\ 
      &   &   &  (keV)     & (10$^{52}$ erg)  &  &  spectrum &  \eiso \\
\hline
      970228 & LONG  &   0.695 &  195$\pm$64 &  1.86$\pm$0.14
 &       SAX &    (1) & (1)  \\
      970508 & LONG  &   0.835 &  145$\pm$43 &  0.71$\pm$0.15
 &       SAX &  (1) & (1)  \\
      970828 & LONG &    0.958 &  586$\pm$117 &  34$\pm$4 & 
      BAT &   (2) & (3)  \\
      971214 & LONG &     3.42 &  685$\pm$133 &  24$\pm$3 & 
      SAX &   (1) & (1)  \\
      980425  & SUB--EN & 0.0085 & 55$\pm$21 & 0.00010$\pm$0.00002 & BAT & 
      (4)  & (4) \\
      980613 & LONG &      1.096 &  194$\pm$89 &  0.68$\pm$0.11
 &       SAX &   (1) & (1) \\
      980703 &  LONG &    0.966 &  503$\pm$64 &  8.3$\pm$0.8
 &       BAT &   (2) & (3) \\
      990123 &  LONG &     1.60 &  1724$\pm$466$^{\rm (b)}$ &  266$\pm$43 & 
      SAX/BAT/KON &       (1,2,5) &  (1,5) \\
      990506 &  LONG &     1.30 &  677$\pm$156$^{\rm (b)}$ &  109$\pm$11 & 
      BAT/KON &       (2,5) & (3,5)  \\
      990510 &  LONG &     1.619 &  423$\pm$42 &  20$\pm$3 & 
      SAX &       (1) & (1)  \\
      990705 &  LONG &    0.842 &  459$\pm$139$^{\rm (b)}$ &  21$\pm$3 & 
      SAX/KON &       (1) &  (1) \\
      990712 &  LONG &    0.434 &  93$\pm$15 &  0.78$\pm$0.15
 &       SAX &   (1) &  (1) \\
      991208 &   LONG &   0.706 &  313$\pm$31 &  25.9$\pm$2.1
 &       KON &   (5) &  (5)  \\
      991216 &   LONG &    1.02 &  648$\pm$134$^{\rm (b)}$ &  78$\pm$8 & 
      BAT/KON &     (2) &  (3) \\
      000131  &   LONG &    4.50 &  987$\pm$ 416$^{\rm (b)}$ &  199$\pm$35 & 
      BAT/KON &    (5) & (5) \\
      000210 &  LONG &    0.846 &  753$\pm$26 &  17.3$\pm$1.9 & 
      KON &    (5) & (5)  \\
      000418 &  LONG &     1.12 &  284$\pm$21 &  10.6$\pm$2.0 & 
      KON &     (5) & (5)  \\
      000911 &   LONG &    1.06 &  1856$\pm$371$^{\rm (c)}$ &  78$\pm$16 & 
      KON &    (6)  & (6) \\
      000926 &   LONG &    2.07 &  310$\pm$20 &  31.4$\pm$6.8 & 
      KON &     (5)  & (5) \\
      010222 &  LONG &     1.48 &  766$\pm$30 &  94$\pm$10 & 
      KON &     (5) & (5)  \\
      010921 &   LONG &   0.450 &  129$\pm$26 &  1.10$\pm$0.11
 &       HET &  (8) & (3) \\
      011211 &   LONG &    2.14 &  186$\pm$24 &  6.3$\pm$0.7
 &       SAX &   (3) & (3)  \\
      020124 &  LONG &     3.20 &  448$\pm$148$^{\rm (b)}$ &  31$\pm$3 & 
      HET/KON &    (8,5) & (3)  \\
      020813 &  LONG &     1.25 &  590$\pm$151$^{\rm (b)}$ &  76$\pm$19$^{\rm (b)}$  & 
      HET/KON &   (5,7) & (3,5)  \\
      020819b &  LONG &    0.410 &  70$\pm$21 &  0.79$\pm$0.20
 &       HET &   (8) & -- \\
      020903 &  XRF &    0.250 &  3.37$\pm$1.79 & 
 0.0028$\pm$0.0007 &       HET &  (9) & (9)  \\
      021004 &   LONG &    2.30 &  266$\pm$117 &  3.8$\pm$0.5
 &       HET &    (8) & (26) \\
      021211  &   LONG &   1.01 &  127$\pm$52$^{\rm (b)}$ & 1.3$\pm$0.15 &
      HET/KON    &    (5,10) & (5,13)  \\
      030226 &   LONG &    1.98 &  289$\pm$66 &  14$\pm$1.5 & 
      HET &  (8) &  (13) \\
      030328 &  LONG &     1.52 &  328$\pm$55 &  43$\pm$4 & 
      HET/KON &  (5,8) & -- \\
      030329 &  LONG &     0.17 &  100$\pm$23$^{\rm (b)}$ & 1.7$\pm$0.3 $^{\rm (b)}$ &
      HET/KON &   (5,8) & -- \\
      030429 &  LONG &     2.65 &  128$\pm$26 &  2.50$\pm$0.30
 &       HET &  (8) & (26) \\
      040924 &   LONG &   0.859 &  102$\pm$35$^{\rm (b)}$ &  1.1$\pm$0.12
 &       HET/KON &   (14,28) & (27)  \\
      041006 &  LONG &    0.716 &  98$\pm$20 &  3.5$\pm$1.0
 &       HET &   (29) & --  \\
      050318* &  LONG &     1.44 &  115$\pm$25 &  2.55$\pm$0.18
 &       SWI &   (16) & (16) \\
      050401* &  LONG &     2.90 &  467$\pm$110 &  41$\pm$8 & 
      KON &  (17) &   (17)   \\
      050416a* & XRF &     0.650 &  25.1$\pm$4.2 &  0.12$\pm$0.02
 &       SWI &  (19) & (19)  \\
      050525* &   LONG &   0.606 &  127$\pm$10 &  3.39$\pm$0.17
 &       SWI & (20) &   --     \\
      050603* &   LONG &    2.821 &  1333$\pm$107 &  70$\pm$5 & 
      KON & (21) &  --  \\
      050709  & SHORT & 0.16 & 100$\pm$16 & 0.0103$\pm$0.0021 & HET & 
      (18) &  (18) \\
      050922c* &  LONG &     2.198 &  415$\pm$111 &  6.1$\pm$2.0 & 
      HET &  (22) &  (22)  \\
      051022* & LONG &  0.80 &  754$\pm$258$^{\rm (b)}$ &  63$\pm$6 & 
      HET/KON &  (23,30) &  --  \\
      051109* & LONG &      2.346 &  539$\pm$200 &  7.5$\pm$0.8
 &       KON &  (24) &  --   \\
      051221* &  SHORT &   0.5465 & 622$\pm$35  &  0.29$\pm$0.06 
 &       KON &  (25) &  (25) \\
\hline
\end{tabular}
\begin{list}{}{}
\item[]Notes. 
$^{\rm (a)}$References for the spectral parameters and for the values and
uncertainties of \epi and \eiso:
(1) \cite{Amati02}, (2) \cite{Jimenez01}, (3) \cite{Amati03},
(4) \cite{Yamazaki03b}, (5) \cite{Ulanov05}, (6) \cite{Price02},
(7) \cite{Barraud03}, (8) \cite{Sakamoto05b}, (9) \cite{Sakamoto04},
(10) \cite{Crew03}, (11) \cite{Atteia03}, (12) \cite{Atteia05},
(13) \cite{Ghirlanda04a}, (14) \cite{Golenetskii04}, (15) \cite{Friedman05},
(16) \cite{Perri05},
(17) \cite{Golenetskii05a},
(18) \cite{Villasenor05},
(19) \cite{Sakamoto05a},
(20) \cite{Cummings05},
(21) \cite{Golenetskii05b},
(22) \cite{Crew05a},
(23) \cite{Golenetskii05c}, 
(24) \cite{Golenetskii05d},
(25) \cite{Golenetskii05e},
(26) \cite{Friedman05},
(27) \cite{Ghirlanda05b},
(28) \cite{Fenimore04},
(29) Official HETE GRB page (http://space.mit.edu/HETE/Bursts/),
(30) \cite{Doty05}
$^{\rm (b)}$ Uncertainties enlarged in order to account for 
significantly different values measured
by different instruments (see text).
$^{\rm (c)}$ Value derived from time resolved spectral analysis by weighting
each time interval with the ratio between its fluence and the total GRB fluence.
A 20\% error is conservatively assumed.
\end{list}
\end{minipage}
\end{table*}

\section{The \epeiso 
correlation}

\subsection{GRBs peak energy and radiated energy} 

The prompt emission spectra of GRBs are non thermal and
in general can be modeled with the Band
function \citep{Band93}, a smoothly broken power--law 
whose parameters are the low energy spectral index, 
$\alpha$, the high
energy spectral index, $\beta$, the break energy, E$_0$, and the overall
normalization. In this model, if $\beta$ $<$ $-$2 then the \nufnu 
spectrum peak energy 
is given by 
\epo
= \eze $\times$ (2 + $\alpha$). 
The spectral shape of most GRBs can be
satisfactorily reproduced by Synchrotron Shock Models (SSM) (e.g. Tavani 
1996 \nocite{Tavani96}): 
the kinetic energy of an
ultra--relativistic fireball (a plasma made of pairs, photons and a small quantity of
baryons) is dissipated into electromagnetic radiation by means of synchrotron
emission originated in internal shocks between colliding shells and/or the external
shock of the fireball with the ISM, see, e.g., \cite{Meszaros02}
and \cite{Piran05} for 
recent reviews.
Nevertheless, the time
resolved analysis of BATSE and \sax GRBs showed that, at least during the initial phase of the
emission, other mechanisms, like Compton up--scattering of UV photons 
surrounding the GRB
source by the ultra relativistic electrons of the fireball or  
thermal
emission by the photosphere of the fireball, may play an important role, see,
e.g., \cite{Preece00,Frontera00b, Ghirlanda03}. The latter emission mechanism
could also be responsible for the smooth curvature characterizing
GRB average spectra and, in particular, may determine the value of \epo
\citep{Ryde05,Rees05}.
A relevant
outcome of the analysis of BATSE events was the evidence of a substantial
clustering of \epo values around 200 keV,
but in the recent years, the discovery and study of X--ray rich events
and X--Ray Flashes (XRFs) 
by \sax and HETE--2, showed that the distribution of
\epo is much less clustered than inferred basing on BATSE data and, in particular,
that it is characterized by a low energy tail extending down at least to $\sim$1 keV
\citep{Kippen01,Sakamoto05b}. 

Since the \sax breakthrough discoveries in 1997, more than
70 redshift estimates have now become available. 
As a
consequence, for these events it is possible to compute the intrinsic
peak energy \epi = \epo $\times (1 + z)$ and the
the radiated energy in a given cosmological 
rest--frame energy band 
following, e.g., the methods described
in \cite{Amati02}, \cite{Amati03} and \cite{Ghirlanda04a}. 
In the simplest assumption of isotropic
emission, the radiated energy, \eiso, ranges from $\sim$10$^{50}$ erg to
$\sim$10$^{54}$ erg for most GRBs and extends down to $\sim$10$^{49}$ erg when
including XRFs \citep{Lamb04,Sakamoto04}. 
When assuming that
the GRB emission is jet--like, based on achromatic breaks observed in the
afterglow decay curves of several GRBs, 
the distribution of the collimation corrected radiated
energy, \ega,  was initially found to be clustered around
$\sim$10$^{51}$ erg \citep{Frail01,Bloom03}; however, recently  
\cite{Ghirlanda04a} showed that, when considering a larger sample of
GRBs with known redshift,
the \ega distribution is broader than inferred before. 

\begin{table*}
\begin{minipage}{155mm}
\caption{\epi and \eiso values for GRBs and XRFs with uncertain estimates, or
upper / lower limits, of 
$z$ or \epo. 
The uncertainties are at 1$\sigma$ significance, 
whereas the upper/lower limits are at 90\% c.l.
The "Type" column
indicates wether the event is a normal long GRB (LONG), an X--Ray Flash
(XRF) or is sub--energetic (SUB--EN).
The "Instruments" column reports the name of the experiment(s), 
or of the satellite(s),
that provided the estimates or
upper/lower limit to the spectral parameters and of the fluence
(BAT = BATSE, SAX = \sax, HET = HETE--2, KON = Konus, SWI = \swift).
The last two columns report the references for the spectral
parameters and the references
for the values and uncertainties (or upper / lower limits)
of \eiso, respectively.
The "--" sign indicates
that \eiso was computed specifically for this work (see text).
GRBs detected by \swift are marked
with an asterisk.}
\begin{tabular}{llllllcc}
\hline
 GRB  & Type & $z$  & \epi  & \eiso & Instruments &  Refs. for $^{\rm (a)}$ &  
 Refs. for $^{\rm (a)}$ \\ 
      &   &   &  (keV)     & (10$^{52}$ erg)  &  &  spectrum &  \eiso \\
\hline
      980326 &  LONG &   1.0 &  71$\pm$36 &  0.56$\pm$0.11
 &       SAX &  (1)  & (1)  \\
      980329 &  LONG &     2.0--3.9 &  935$\pm$150 &  177$\pm$58$^{\rm (b)}$ & 
      SAX &    (1) & (1,12)  \\
      981226 &  LONG &     1.11 &  $<$160$^{\rm (c)}$ &  0.59$\pm$0.12
 &       SAX &  (13) & -- \\
      000214 & LONG & 0.37--0.47 &  $>$117 & 
 0.93$\pm$ 0.03 &       SAX &    (1) & (1)  \\
      001109 &  LONG &    0.40 &  101$\pm$45 &  0.40$\pm$0.02
 &       SAX &  (3) & --  \\
      011121 &  LONG &    0.360 &  793$\pm$533$^{\rm (b)}$ &  9.9$\pm$2.2$^{\rm (b)}$ & 
      SAX/KON &     (7,11) & (7,11)  \\
      020405 &  LONG &    0.69 &  612$\pm$122 &  12.8$\pm$1.5 & 
      SAX &   (4)  & (5)  \\
      030323 &   LONG &    3.37 &  270$\pm$113 &  3.2$\pm$1.0
 &       HET &  (12) & (27)  \\
      030723 & XRF & $<$2.3 & $<$0.023 & $<$16. & HET & (6) & (5)  \\
      031203 &  SUB--EN &    0.105 &  158$\pm$51 & 
 0.010$\pm$ 0.004 &      KON  &   (7) & (7,14)  \\
      050315* & LONG & 1.949 & $<$89 & 4.9$\pm$1.5 & SWI & (8) & (8)  \\
      050824* &  LONG &    0.83 &  $<$23 & 
 0.130$\pm$ 0.029 &   HET & (9) & --   \\
      050904* &   LONG &    6.29 &  $>$1100 &  193$\pm$127
 &       SWI &  (10) & (10)  \\
\hline
\end{tabular}
\begin{list}{}{}
\item[]Notes. 
$^{\rm (a)}$References for the spectral parameters and for the values and
uncertainties of \epi and \eiso:
(1) \cite{Amati02}, (2) \cite{Christensen05}, (3) \cite{Amati03b},
(4) \cite{Price03},
(5) \cite{Ghirlanda04a}, (6) \cite{Sakamoto05b}, (7) \cite{Ulanov05},
(8) \cite{Vaughan06},
(9) \cite{Crew05b},
(10) \cite{Cusumano06},
(11) \cite{Amati03},
(12) \cite{Jimenez01},
(13) \cite{Frontera00a},
(14) \cite{Sazonov04}
$^{\rm (b)}$ Uncertainties enlarged in order to account for 
significantly different values measured
by different instruments (see text).
$^{\rm (c)}$ Upper limit inferred from time resolved spectral analysis. 
\end{list}
\end{minipage}
\end{table*}

\subsection{Discovery, confirmation and extension of the
\epeiso correlation}

In 2002, \cite{Amati02} presented the results of the analysis of 
the average 2--700 keV
spectra of 12 \sax GRBs with known redshift (9 firm
measurements and 3 possible values). 
The more relevant outcome of this
work was
the evidence of a strong correlation between \epi and \eiso .  
The linear correlation coefficient between log(\epi) and log(\eiso)  was found to
be 0.949 for the 9 GRBs with firm redshift estimates, corresponding to a chance
probability of $\sim$0.005\%.  
The slope of the power--law best describing the trend of
\epi as a function of \eiso was $\sim$0.5.  This work was extended by
\cite{Amati03} by including in the sample 10 more events with known redshift
for which new spectral data (\sax
events) or published best fit spectral parameters (BATSE and HETE--2 events) were
available. The \epeiso correlation was confirmed and its significance
increased, giving a correlation coefficient similar to that derived by \cite{Amati02}
but with a much higher number of events. Basing on HETE--2 measurements,
\cite{Lamb04}, \cite{Sakamoto04} and \cite{Sakamoto05a}
not only confirmed the \epeiso correlation 
but remarkably
extended it to XRFs, showing that it holds over three orders of magnitude in \epi
and five orders of magnitude in \eiso . 
The addition of new data, as more
redshift estimates became available, confirmed the correlation and increased
its significance, as found e.g. by \cite{Ghirlanda04a} (29 events, chance 
probability of 7.6$\times$10$^{-7}$). 
Finally, the relevance of the \epeiso correlation
for the GRB field stimulated several similar studies, which led to
the discovery of
correlations of \epi with other
GRB intensity indicators like the average isotropic equivalent luminosity $L_{\rm iso}$
\citep{Lamb04,Lamb05} and the peak isotropic
equivalent luminosity $L_{\rm p,iso}$ \citep{Yonetoku04,Ghirlanda05a}.
\cite{Liang04} also showed that, at least for a good fraction of events,
the \epi -- $L_{\rm iso}$ correlation holds also within GRBs.
All these correlations show the same slope and dipersion of the \epeiso correlation,
and reflect the tight correlation existing
between \eiso, $L_{\rm iso}$ and $L_{p,\rm iso}$ \citep{Lamb04,Lamb05,Ghirlanda05a}. 
More intriguing, as will be discussed in Section 5, are the 
\epega and \epeisotb ($t_b$ is the achromatic afterglow light curve break time) 
correlations discovered by \cite{Ghirlanda04a} and
\cite{Liang05} basing on a limited sample of GRBs with known $z$, \epi and $t_b$.
Also, evidence of a strong correlation between \epi, \eiso
and the "high--signal" time scale $T_{0.45}$ has been found very
recently \citep{Firmani06}.

The outcome of the analysis of the \epeiso correlation performed in previous
works are summarized in the first five lines of Table 3. As it can be seen,
the chance probability of this correlation is very low and decreases when 
increasing the number of events in the sample considered. Nevertheless,
the fits with a power--law are always very poor (as indicated by the reported
\cqr values) and both the normalization
and the index vary significantly depending on the sample considered. This is
an effect of the extra--Poissonian dispersion of the correlation, as 
will
be discussed in detail in the next Section.

\section{Updated sample and distributions of \epi and \eiso .}
The last three lines of Table 3
report the
results of the analysis that I performed on the most updated (as of
December 2005)
sample of long
GRBs/XRFs 
with firm estimates 
of both $z$ and \epo . 
This sample, reported in Table 1, consists of a total of 41 events and includes
events already considered in previous
works, new events, such as \swift GRBs, and older events 
for which useful spectral information has become available only
recently, as is the case e.g. for some events detected by Konus/Wind
\citep{Ulanov05}. 
Table 1 includes also the peculiar 
sub--energetic avent
GRB980425 and the only two short GRBs (GRB050709 and GRB051221) 
for
which firm estimates of $z$ and \epo are available.
Table 2 includes 12 classical long GRBs and the other sub--energetic
event GRB031203, with uncertain
estimates of $z$ or \epo. It also includes XRF030723, for which only an 
upper limit to the redshift is available. 
The \epi and \eiso values of more than half of the events are
taken from \cite{Amati02}, \cite{Amati03},\cite{Ghirlanda04a},\cite{Friedman05},
\cite{Ulanov05},
while for the other events they are taken from specific references,
mostly GCNs. 
For HETE--2 GRBs with estimates of \epo available
from both \cite{Barraud03}
and \cite{Sakamoto05a}, I used the values from \cite{Sakamoto05a}, which are based
on joint WXM/FREGATE analysis. The only exception to this rule is GRB020813, for which
a depletion of X--ray photons in the WXM energy band was observed by means of
time resolved spectral analysis \citep{Sato05a}.
Some GRBs included in the samples of some of the works mentioned above, were 
excluded from the sample of  GRBs with firm estimates of $z$ and \epi because of a too large difference between the \epo measured by different detectors (e.g. GRB011121) or
a poorly unconstrained 90\% c.l. interval of \epo (e.g. GRB030323) (see references
in Table 2).
For all GRBs, the
last two columns of Tables 1 and 2 report the references for the spectral parameters
and for the values and uncertainties of \epi and \eiso, including also the
methods adopted to derive these quantities based on the measured
spectra and fluences. 
It has to be noted that in the computation of \eiso some authors
assume $H_0$=65 km s$^{-1}$ Mpc$^{-1}$ (e.g., Amati et al. 2002), while others assume
$H_0$=70 km s$^{-1}$ Mpc$^{-1}$ (e.g., Ghirlanda, Ghisellini \& Lazzati 2004);
instead, the values of  $\Omega_m$ and  $\Omega_\Lambda$ are always assumed to be
0.3 and 0.7, respectively.
Thus, in order to have a homogeneous data set, I corrected the \eiso values
computed by assuming $H_0$=70 km s$^{-1}$ Mpc$^{-1}$ by accounting for the
different luminosity distance obtained with $H_0$=65 km s$^{-1}$ Mpc$^{-1}$. 

The \eiso values
from \cite{Ulanov05} where computed by integrating the observed fluence between
10/(1+z) and 10000/(1+z) keV (Ulanov, private communication) and thus 
between 10--10000 keV in the GRB cosmological
rest-frame . The difference with respect to integrating
from 1 to 10000 keV clearly depends on the spectral
parameters, in particular on $\alpha$, and 
is not higher than 10\% in the
worst cases (and typically of the order of 3-5\%).
Given that the values of $\alpha$ are not reported by \cite{Ulanov05}, I
corrected their \eiso values and 1$\sigma$ confidence intervals 
by conservatively assuming that the extension of the integration from 10
keV to 1 keV my contribute from 0 to 10\% to the total radiated energy.

For some events of the sample, marked with
a note in Tables 1 and 2, I enlarged the uncertainties on \epi and/or \eiso 
in order to include the different central values and 1$\sigma$ uncertainties
reported in the literature
based on measurements by different GRB detectors. 
The issue of the systematics
in the estimates of spectral parameters and fluence 
due to detectors limited energy bands and 
sensitivities 
as a function of photon energy is discussed in Sections 5.3 and 6.2 .
For those few events for which there is no reference 
directly reporting the values of \epi and/or \eiso, these have been calculated
based on published spectral parameters and fluence by following the methods mentioned 
in previous Section and detailed e.g. in  \cite{Amati02} and \cite{Ghirlanda04a}.
In particular, the unpublished values of \eiso have been computed in the
1--10000 keV cosmological rest--frame energy band and by assuming a standard
cosmology with $\Omega_m$=0.3, $\Omega_\Lambda$=0.7 and 
$H_0$=65 km s$^{-1}$ Mpc$^{-1}$. For these events, the uncertainties on \epi and
\eiso were computed by propagating the fractional errors on \eze and fluence, respectively,
as done, e.g., in \cite{Amati03} for BATSE and HETE--2 events. 
The values of $z$ have been taken directly from J. Greiner's  GRB Table\footnote{ 
http://www.mpe.mpg.de/$\sim$jcg/grbgen.html}, which also includes complete references.
For those cases in which only a range for $z$ is available
(Table 2), \epi and \eiso have been computed by assuming the central value.
In the case of XRF030723 (Table 2), the upper limit to the redshift comes from
\cite{Fynbo04} and the upper limits to \epi and \eiso from \cite{Lamb04}.

It is important to notice that GRBs included in these samples have been detected,
and their spectral parameters measured, by detectors with different sensitivities 
and energy bands. This, together with the fact that \swift is allowing $z$ determinations
for more types of GRBs, should reduce significantly the possible impact of selection effects
. This issue will be discussed in Section 6.\\
In Figure 1 I show the logarithmic distributions of \eiso (left panel) and \epi
(right panel). These refer to all the GRBs included in Table 1 plus GRB031203
(Table 2). As can be seen, in both cases 
the bulk of the distribution can be fitted by a Gaussian, but a low energy tail
is evident. In the log(\eiso) 
distribution, the low tail is due to the sub--energetic GRBs 980425 and 031203, 
the XRF 020903 and the short GRB 050709;
the fit of this distribution with a Gaussian gives an average of 
$\sim$10$^{53}$ erg
and a logarithmic dispersion of $\sim$0.9 . It is noticeable that the \eiso distribution 
spans about 6 orders of magnitude. The fit of the log(\epi) 
distribution with a Gaussian
gives an average of $\sim$350 keV and a logarithmic dispersion of $\sim$0.45 . In 
this case the low--energy tail is given by the XRFs 020903 and 050416a, 
while sub--energetic and short GRBs show \epi values consistent with the bulk of
the distribution.
It is worthing to note that the \epi distribution is much broader than 
the \epo distribution inferred from BATSE GRBs
\citep{Kaneko06}.

\begin{figure*}
\centerline{\includegraphics[width=9cm]{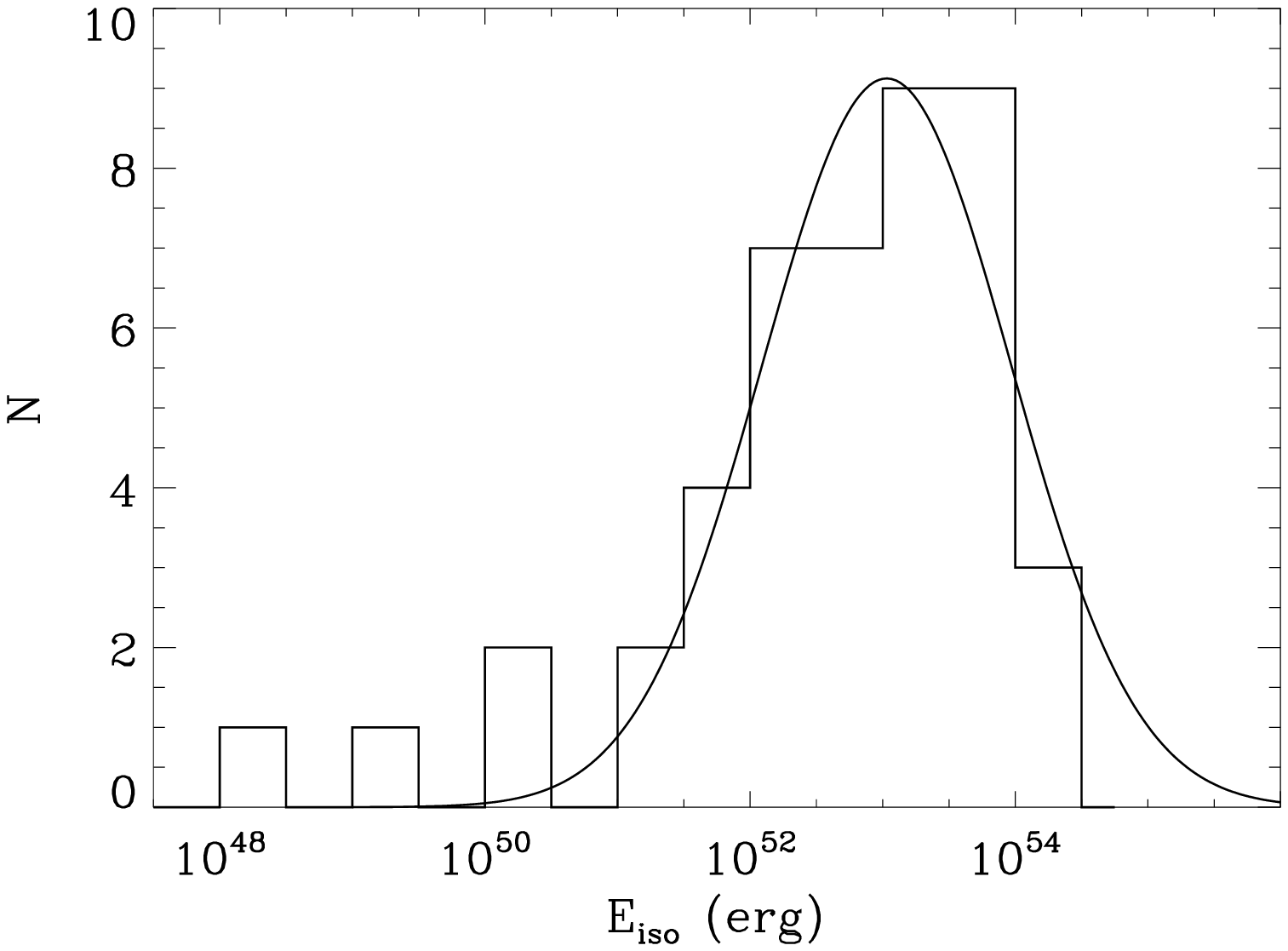} 
\includegraphics[width=9cm]{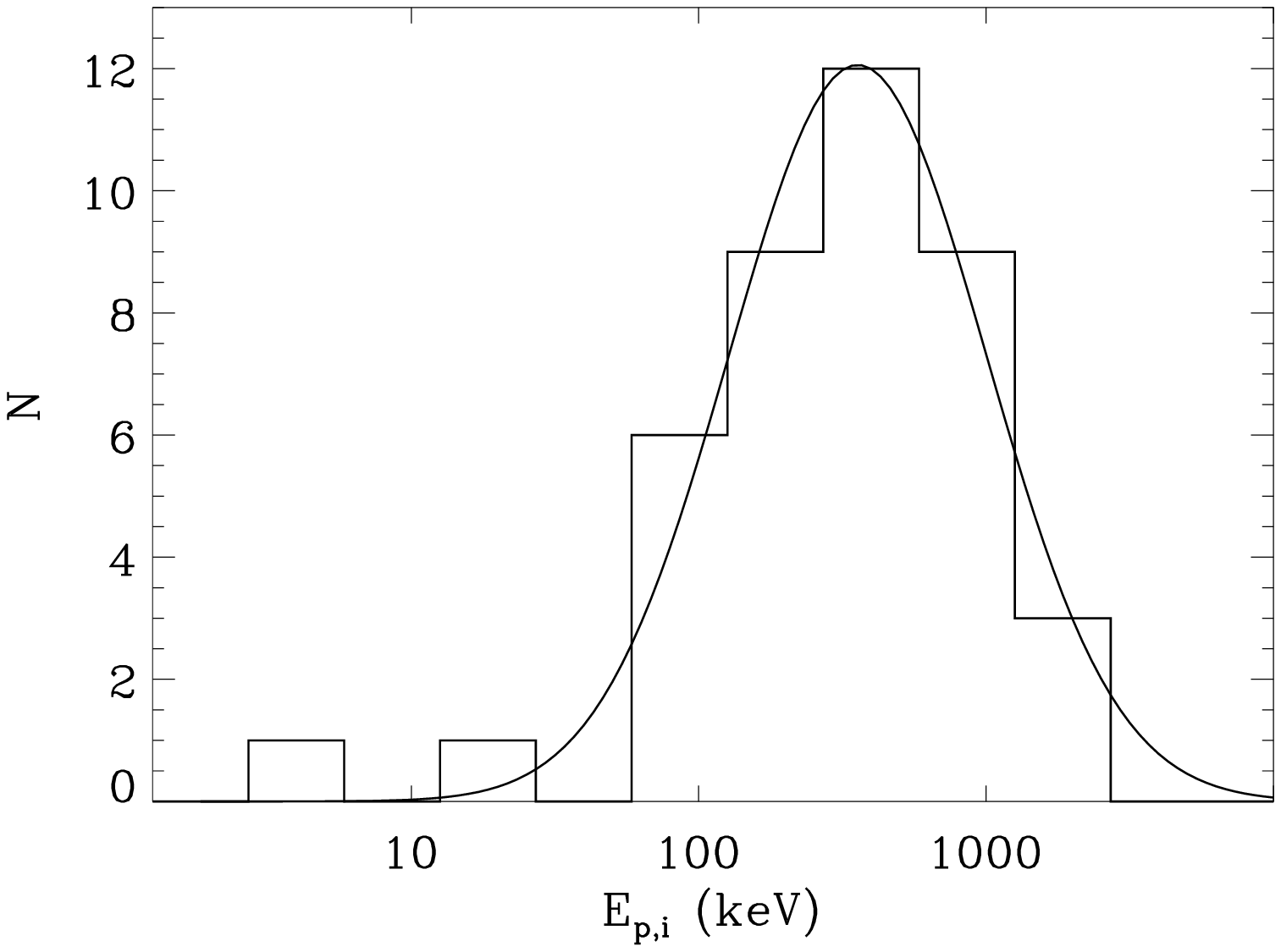}}
\caption{Distributions of log(\eiso) (left) and log(\epi) (right)
for the 41 GRBs with firm redshift and \epo, 
the short GRBs 
050709 and 051221 and the sub--energetic GRBs 980425 and 031203.
For both distributions, the best fit Gaussian is superimposed to the data.
}
\end{figure*}

\section{The \epeiso correlation: re--analysis}

The (\epi,\eiso) points corresponding to the 41 GRBs/XRFs with with firm estimates
of $z$ and \epi, all included in Table 1, are shown in Figure 2, whereas the 
(\epi,\eiso) points and upper/lower limits corresponding to the GRBs with 
uncertain $z$ and \epi (Table 2) are shown in Figure 3, which also includes the points 
corresponding to the peculiar sub--energetic GRB980425 and the two short GRBs 
050709 and 051221 . In both Figures, the point corresponding to \swift GRBs are
shown as filled circles. The first 
two lines of the second part of Table 3 report the results of the 
analysis performed on the sample plotted in Figure 2.
The correlation
analysis is based on the estimate of the Spearman's rank correlation coefficient
between \epi and \eiso and the fits  with a
power--law \epi = K $\times$ $E_{iso}^m$ are performed by accounting for
the errors on both \epi and \eiso .
As can be seen, with the updated sample subject of this analysis, 
the chance probability of correlation between  the logarithms of \epi and \eiso
is as low as $\sim$3.5$\times$10$^{-14}$, when considering only the 41 
classical GRBs, and
$\sim$10$^{-14}$ when including also the XRFs 020903 and 050416a.
Thus, increasing the sample by adding new data and
making it more complete (as mentioned above) not only confirms the \epeiso 
correlation but also
reduces its chance probability to a negligible value. The index of the power--law,
$\sim$0.57, 
is found to lie in the range 0.4--0.6, consistently with the findings of  
previous analysis,
and does not change significantly by
including or not in the sample XRFs 020903 and 050416a (see Table 3). 
The value of the normalization
is found to be somewhat lower with respect to previous analysis, except
for the recent analysis performed by Nava et al. (2006) on a subsample of 18 GRBs
which includes more recent events with respect to previous works. The
power--law best fitting the data of the 41 GRBs with firm estimates of $z$ and \epi
is shown as a dashed line in Figures 2 and 3.

\begin{table*}
\begin{minipage}{155mm}
\caption{Summary of the results of the analysis 
of the \epeiso 
correlation as reported in previous works (top) and performed in this work
(bottom). 
The coefficient $\rho$ is the Spearman's rank correlation coefficient
between $E_{p,i}$ and $E_{iso}$. N is the number of events considered, 
m, K and $\chi$$^2_\nu$ refer to fits of the
\epeiso correlation with a power--law \epi = K$\times$$E_{iso}^m$  
(\epi is in keV and \eiso in units of 10$^{52}$ erg).
The 
uncertainties reported in the first part of the Table (values taken from the literature) are at 1$\sigma$ confidence level, while those reported in the second part (results obtained
in this work) are at 90\% significance.
When not available in the literature, the values of K and $\chi$$^2_\nu$  reported
in the top part of the Table
have been computed specifically for this work.} 
\begin{tabular}{llclccc}
\hline
 Reference & N & $\rho$ & Chance Prob.  & m & K & $\chi$$^2_\nu$ \\
\hline
Amati et al. (2002) &  9  & 0.92 & 5.0$\times$10$^{-4}$ & 0.52$\pm$0.06 & 105$\pm$11 & 3.9\\
Amati  (2003) &  20  & 0.92 & 1.1$\times$10$^{-8}$ & 0.35$\pm$0.06 & 118$\pm$9 & 6.1 \\
Ghirlanda, Ghisellini \& Lazzati (2004) &  27
 & 0.80 & 7.6$\times$10$^{-7}$ & 0.40$\pm$0.05 & 95$\pm$7 & 6.2 \\
Friedman \& Bloom (2005) & 29 & 0.88 & 4.9$\times$10$^{-10}$ & 0.50$\pm$0.04 & 90$\pm$8 & 9.5 \\
Nava et al. (2006) & 18 & 0.82 & 3.1$\times$10$^{-5}$ & 0.57$\pm$0.02 & 71$\pm$2 & 5.2 \\
\hline
This work (only GRBs) & 39 & 0.89 & 3.1$\times$10$^{-14}$ & 0.57$_{-0.02}^{+0.02}$ & 80$_{-4}^{+4}$ & 7.2 \\
This work (including XRFs 020903 and 050416) & 41 & 0.89 & 7.0$\times$10$^{-15}$ & 0.57$_{-0.02}^{+0.02}$ & 81$_{-2}^{+2}$ & 6.9 \\
This work (accounting for sample variance) & 41 & 0.89 &  7.0$\times$10$^{-15}$ & 
0.49$_{-0.05}^{+0.06}$  
& 95$_{-9}^{+11}$ & 0.95  \\
\hline
\end{tabular}
\end{minipage}
\end{table*}

Despite the correlation is very highly significant,
the $\chi^2$ values obtained by fitting the data
with a power--law are poor, as found in previous analysis (Table 3). This means that the
scatter of the data around the best fit model cannot be due only to statistical
fluctuations, unless the systematics in the estimates of \epi and \eiso are strongly
under--estimated, as will be discussed in next Section. Another effect of the dispersion
characterizing the correlation is that, 
as mentioned above, the slope and normalization of the power--law are found 
to change significantly depending on the sub--sample considered. This 
extra--Poissonian dispersion of the \epeiso correlation,
which potentially contains precious
information (as will be discussed in Section 5) and has to be taken into account
when testing it (as will be discussed in Section 6), 
can be quantified by introducing
in the modelization a further parameter commonly called
"sample variance" or "slop". 
The issue of fitting data
with a power--law by accounting simultaneously for X and Y errors and for 
sample variance has been faced e.g. by \cite{Reichart01} and 
\cite{Reichart05}
when analyzing the 
peak luminosity--variability correlation in GRBs.

The methods used in these works are based on a likelihood function derived
with a bayesian approach to the problem; however, recently \cite{Dagostini05}
and \cite{Guidorzi06}
showed that the correct likelihood function is slightly different from
that used by \cite{Reichart01,Reichart05}.
I applied the method by \cite{Dagostini05} and \cite{Guidorzi06}
to the sample of 41 GRBs
considered above; with this modelization the parameters are the index
and normalization of the power--law ($m$ and $K$) and the logarithmic
dispersion of \epi ($\sigma$$_{logE_{p,i}}$).
The result of this analysis is reported in the last line of Table 3. 
The values of the index and normalization of the best--fit
power--law, $\sim$0.5 and $\sim$100, respectively, lie
in the ranges of values found with different samples by adopting
the simple power--law fit and are coincident with those usually assumed 
in the literature when comparing new data with the \epeiso correlation or
using it as an input or required output for GRBs/XRFs  synthesis models
(see next Section), basing on the early results from \cite{Amati02} (
Table 1, first line). 
The best fit power--law obtained with this method is plotted in Figures
2 and 3 as a continuous line.
The value of the sample variance resulting from the fit is
$\sigma$$_{logE_{p,i}}$ = 0.15$_{-0.04}^{+0.04}$ (90\% c.l.);
in Figures 1 and 2 I show the region,
delimitated by two dotted lines, corresponding to deviations of \epi from
the best fit power--law of $\sim$2.5$\sigma$$_{logE_{p,i}}$, assuming 
$\sigma$$_{logE_{p,i}}$=0.15 (the central value of the 90\% confidence interval).
For a comparison, as shown in Figure 4, the dispersion of the log(\epi) 
central values around the best fit power--law obtained without accounting
for sample variance (line 7 of Table 3)
can be fitted with a Gaussian
with dispersion $\sim$0.21, consistently with previous analysis based on
smaller samples. A similar value ($\sigma$$_{logE_{p,i}}$$\sim$0.2, is
obatined when computing the scatter of the central data points around the law
\epi = 95 $\times$ $E_{iso}^{0.49}$ .
This value is of course higher than sample
variance,
because it includes statistical fluctuations; if one considers this as the
overall dispersion of the correlation, then the two dotted lines in Figures 2 and 3 
delimitate
the $\pm$2$\sigma$ region. An estimate of the sample variance characterizing
the \epeiso correlation was also performed by \cite{Lamb05}, who, based
on a smaller sample and following a different method, derived a value of
$\sim$0.13 .\\
Finally, from Figure 3, it can be seen that the uncertain values and upper/lower limits
of \epi and \eiso of classical long GRBs and XRFs (reported in Table 2)
are consistent with the \epeiso
correlation, including \swift GRBs. Figure 3 also shows clearly that the
\epeiso plane can be very useful in discriminating different classes of
GRBs. Indeed, both sub--energetic GRBs (GRB980425 and, possibly, GRB031203)
and short GRBs (050709 and 051221) are clear outliers of the correlations, 
showing \eiso values
too low with respect to their peak energies, which range within those of
normal GRBs (Figure 1, right panel).

During the reviewing process of this article, firm estimates of both $z$ and \epo
have become available for 4 more Swift events (GRB060115, GRB060124, GRB060206, 
GRB060418), based on spectral measurements from
Swift, HETE--2, Konus. The \epi and \eiso of these events are all
fully consistent with the \epeiso correlation.

\begin{figure*}
\centerline{\includegraphics[width=12cm]{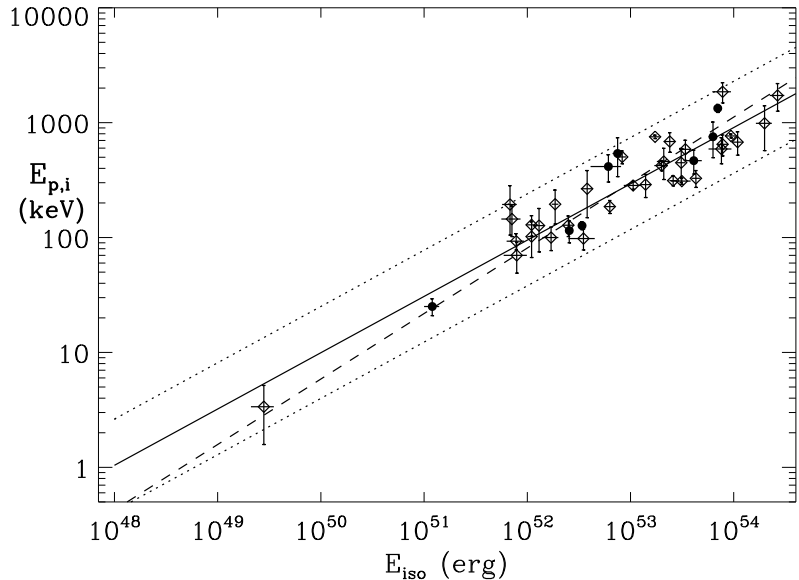}}
\caption{\epi and \eiso values for 41 GRBs/XRFs with firm 
redshift and \epo 
estimates. Filled circles correspond to \swift GRBs.
The continuous line is the best fit power--law \epi = 95 $\times$ $E_{\rm iso}^{0.49}$
obtained by accounting for sample variance; the
dotted lines delimitate the region corresponding to a vertical logarithmic deviation of 0.4.
The dashed line is the best fit power--law \epi = 77 $\times$ $E_{\rm iso}^{0.57}$
obtained by fitting the data without accounting for sample variance. See text
for details.}
\end{figure*}

\begin{figure*}
\centerline{\includegraphics[width=12cm]{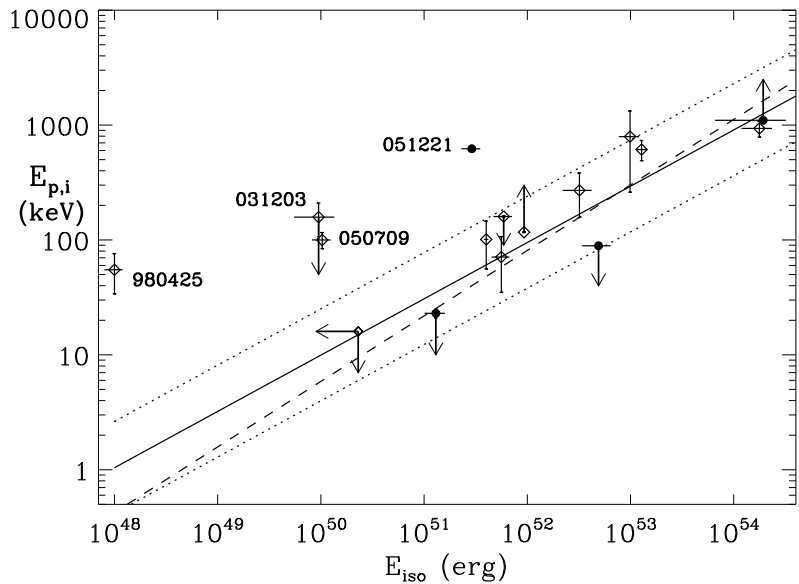}}
\caption{Same as Figure 2 for the 12 GRBs with uncertain estimates
of $z$ and/or \epo , for the peculiar sub--energetic event GRB980425 / SN1998bw
and for
the two short GRBs 050709 and 051221.}
\end{figure*}

\section{Main implications and discussion}

The analysis presented in previous Sections, based on an updated sample containing
about twice events with respect to previous works and including
the recent \swift GRBs,
confirms and strenghtens the \epeiso
correlation for long GRBs/XRFs and gives its best characterization 
up to now in terms of index and
normalization of the best--fit power--law and of its dispersion. 
Remarkably, the correlation now extends over $\sim$5 orders of magnitude
in \eiso,  $\sim$3 orders of magnitude in \epi and over a redshift range
$\sim$0.15$<$$z$$<$4.5 (but also the highest $z$ event, GRB050904 at $z$=6.29,
has \epi and \eiso values consistent with it).
Since its discovery in 2002 \citep{Amati02} and in particular its confirmation
and extension to XRFs \citep{Amati03, Lamb04, Sakamoto04}, the origin
of the \epeiso correlation and its implications for GRB models have been
investigated by several works. The impact of this robust observational
evidence on prompt emission models concerns mainly the physics, the geometry
(i.e. shape and properties of jets),
viewing angle effects and GRB/XRF unification. 
Indeed, the existence 
of the \epeiso correlation and its properties are also often used 
as an ingredient or a test
output for
GRB synthesis models, as in the case of, e.g., the GRB/XRF model by 
\cite{Barraud05}, 
the multi--subjets model by \cite{Toma05}, the uniform jet model 
by \cite{Lamb05}, the study of the impact of off--jet relativistic kinematics
by \cite{Donaghy05a}, the dissipative photosphere models by \cite{Rees05},
the supercritical pile model by \cite{Mastichiadis06}.
Another important outcome of the analysis presented in this paper is
the clear evidence that, in addition to the peculiar sub--energetic GRB980425
(and possibly GRB031203),
short GRBs do not follow the \epeiso 
correlation, as suggested by the different location between
long and short BATSE GRBs in the hardness--intensity plane.
Below I summarize these topics and discuss also the possible origin of the
extra--Poissonian dispersion of the correlation. 

\subsection{Physics of prompt emission}

The physics of the prompt emission of GRBs is still far to
be settled and a variety of scenarios, within the standard fireball picture, have
been proposed, based on different emission mechanisms
(e.g. SSM internal shocks, Inverse Compton dominated internal
shocks, SSM external shocks, photospheric emission dominated models) and different
kinds of fireball (e.g. kinetic 
energy dominated
or Poynting flux dominated), see e.g. \cite{Zhang02} for a review. 
In general, both \epi and \eiso are linked to the fireball bulk Lorentz factor,
$\Gamma$, in a way that varies in each scenario, and the existence and
properties of the \epeiso correlation
allow to constrain the range of values 
of the parameters, see, e.g., \cite{Zhang02} and \cite{Schaefer03}. 
For instance, as shown, e.g., by \cite{Zhang02}, \cite{Ryde05} and
\cite{Rees05}, 
for a power--law electron
distribution generated in an internal shock within a fireball with bulk
Lorentz factor $\Gamma$, it is possible to derive the relation
\epi $\propto$ $\Gamma$$^{-2} L^{1/2} t_\nu^{-1}$ , where $L$ is the GRB luminosity and
$t_\nu$ is
the typical variability time scale. Clearly, in order to produce the observed
\epeiso correlation the above formula would require that $\Gamma$ and $t_\nu$ 
are approximately the same for all GRBs, an assumption which is difficult to 
justify. Things get even more complicated if one
takes into account that the models generally assume $L$ $\propto$ $\Gamma^{\beta}$,
with the value of $\beta$ varying in each scenario and 
is typically $\sim$2--3 \citep{Zhang02,Schaefer03,Ramirez05a,Ryde05}.
More specific examples of the constraints put by the \epeiso 
correlation on the parameters
of SSM and IC based
emission models, both in internal and external shocks, can be found, e.g.,in 
\cite{Zhang02,Schaefer03}.
An interesting possibility, which is currently the subject of many 
theoretical works, is that a substantial contribution to 
prompt radiation of GRBs comes
from direct or Comptonized thermal emission from the photosphere 
of the fireball \citep{Zhang02,Ryde05,Rees05,Ramirez05a}.
This could explain the very hard spectra observed at the beginning of several 
events \citep{Preece00,Frontera00b,Ghirlanda03}, inconsistent with 
SSM models, 
and the smooth curvature characterizing 
GRBs average spectra.
In this scenarios, \epi is mainly determined by the peak temperature $T_{pk}$
of black--body distributed photons
and thus naturally linked to the luminosity or radiated 
energy. For instance, for Comptonized emission from the photosphere 
one can derive the
relations \epi $\propto$ $\Gamma$T$_{pk}$ $\propto$ $\Gamma$$^{2} L^{-1/4}$ 
or \epi $\propto$ $\Gamma$T$_{pk}$ $\propto$ r$_0$$^{-1/2} L^{1/4}$ 
(where $r_0$ is a particular distance between the central engine and the emitting
region),
depending on the assumptions made
\citep{Rees05}. Also in this case, off course, the $L$ $\propto$ $\Gamma^{\beta}$
relation plays a decisive role. As shown by \cite{Rees05}, in this scenario 
the correct
\epeiso relation can be obtained for some specific physical conditions
just below the photosphere. 

Finally, also, the fact that the \epi
distribution is broader than inferred previously basing mainly on the observed \epo values
of bright BATSE GRBs, as shown in Figure 1, can put important constraints on
models for the physics of GRB prompt emission
\citep{Zhang02}.

\begin{figure}
\includegraphics[width=9cm]{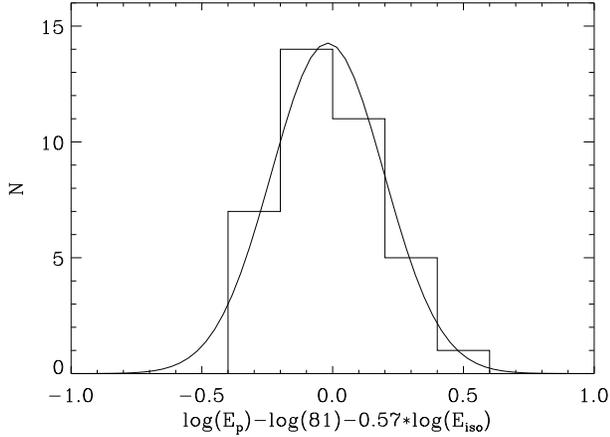}
\caption{Logarithmic dispersion of the \epi values of the 41 GRBs with firm redshift and \epi 
estimates around the power--law best fitting the \epeiso correlation without 
accounting for sample variance. I also show the 
best fit Gaussian, which has a dispersion of $\sigma_{log(E_p,i)}$ $\sim$ 0.21 .}
\end{figure}

\subsection{Jets, viewing angle effects and GRB/XRF unification models}

The validity of the \epeiso correlation from the
most energetic GRBs to XRFs (see Figure 2)
confirms that these two phenomena have the same origin
and is a very challenging observable for GRB jet models. Indeed, these models have to
explain not only how \eiso and \epi are linked to the jet opening angle,
$\theta$$_{jet}$, and/or to the viewing angle with respect to the jet axis,
$\theta$$_v$, but also how \eiso can span over several orders of magnitudes. In
the most simple scenario, the uniform jet model \citep{Frail01,Lamb05}, 
jet opening angles are variable
and the observer measures the same value
of \eiso independently of $\theta$$_v$. In the other popular scenario, the
universal structured jet model (e.g. Rossi, Lazzati \& Rees \nocite{Rossi02}), 
\eiso depends on $\theta$$_v$. 
As discussed in
Section 2, in the hypothesis that achromatic breaks found in the afterglow light curves
of some GRBs with known redshift are due to collimated emission, it was originally
found \citep{Frail01,Berger03} that the
collimation corrected radiated energy, \ega, is of the same order 
($\sim$10$^{51}$ erg) for most GRBs and that
\eiso $\propto$ $\theta$$_{jet}^{-2}$, assuming a uniform jet. In the case of
structured jet models, which assume that $\theta$$_{jet}$ is similar for all GRBs 
(hence this scenario is also called universal jet model) the same observations imply 
that \eiso $\propto$
$\theta$$_{v}^{-2}$. Thus, always under the assumption of a nearly constant
\ega, the found \epeiso correlation implies
\epi $\propto$ $\theta$$_{jet}^{-1}$ and \epi $\propto$ $\theta$$_{v}^{-1}$
for the uniform and structured jet models, respectively. \cite{Lamb05} argue
that the structured universal jet model, in order to explain the validity of the
\epeiso correlation from XRFs to energetic GRBs, predicts a number of
detected XRFs several orders of magnitude higher than the observed one ($\sim$1/3
than that of GRBs). In their view, the uniform jet model can overcome these problems
by assuming a distribution of jet opening angles N($\theta$$_{jet}$) $\propto$
$\theta$$_{jet}^{-2}$. This implies that the great majority of GRBs have opening
angles smaller than $\sim$1$^{\circ}$ and that the true rate of GRBs is several
orders of magnitude higher than observed and comparable to that of SN Ic. On the
other hand, \cite{Zhang04} show that the requirement that most GRBs have jet opening
angles less than 1 degree, needed in the uniform jet scenario in order to explain the
\epeiso correlation, as discussed above, implies values of the
fireball kinetic energy and/or of the interstellar medium density much higher than
those inferred from the afterglow decay light curves. Together with other authors,
e.g. \cite{Lloyd04,Dai05}, they propose a modification of the universal structured jet
model, the quasi--universal Gaussian structured jet.  In this model, the measured
\eiso undergoes a mild variation for values of $\theta$$_{v}$ inside a typical
angle, which has a quasi--universal value for all GRBs/XRFs, whereas it decreases
very rapidly (e.g. exponentially) for values outside the typical angle. In this way,
the universal structured jet scenario can reproduce the \epeiso
correlation and predict the observed ratio between the number of XRFs and that of
GRBs. Recently, a Fisher--shape has been proposed, for both the variable
angle and universal angle scenarios \citep{Donaghy05b,Dai05},
as a very promising alternative, in particular for the explanation
of the validity of the \epeiso correlation from the brightest
GRBs to XRFs. Other jet models proposed very recently that can reproduce
the \epeiso correlation  
include the ring--shaped jet model, see \cite{Eichler04}, and the multi--component
(subjets) model, see \cite{Toma05}.

Of particular interest are the off--axis scenarios, in which the jet is typically
assumed to be
uniform but, due to relativistic beaming and Doppler
effects, for $\theta$$_{v}$ $>$
$\theta$$_{jet}$ the measured emissivity does not sharply go to zero 
and the event is detected by the observer with \eiso and \epi dropping
rapidly as $\theta$$_{v}$ increases 
\citep{Yamazaki03a,Granot02,Eichler04,Donaghy05a}. 
In these models, XRFs are those events seen very
off--axis and the XRFs rate with respect to GRBs and the \epeiso
correlation can be correctly predicted. As shown 
e.g. in \cite{Yamazaki04} for a simple model of GRB jet,
if the Doppler shift factor is 
$\delta = [\Gamma(1 - \beta cos(\theta_{v} - \theta_{jet})]^{-1}$ (where $\beta$ 
is the
velocity of the outflow in units of speed of light),
\epi and \eiso scale, with respect to their values observable at the edge
of the jet, as \epi $\propto$ $\delta$ and \eiso $\propto$ $\delta^{1-\alpha}$ ,
where $\alpha$ is the spectral index of the prompt emission photon spectrum in 
the hard
X--ray energy band. By combining these relations one can obtain the \epeiso
correlation with index 0.5 for classical GRBs ($\alpha$$\sim$$-$1) and 0.3
for XRFs ($\alpha$$\sim$$-$2). A detailed study of off--jet relativistic
kinematics effects has been recently performed by \cite{Donaghy05a} for a set
uniform (i.e. top hat shaped -- variable opening angle) jet models, finding
that these scenarios predict a significant population of bursts away from
the \epeiso correlation, unless $\Gamma$ $>$ 300 for all bursts or there is
a strong anti--correlation between $\Gamma$ and the jet solid angle.
Finally, the off--axis effects for very weak and soft events can be applied 
in a similar way as described above in the context of the cannon ball
(CB) model for GRBs in order to reproduce the \epeiso correlation
\citep{Dar04}. 

\subsection{The dispersion of the correlation}

In addition to its existence and slope, also the extra--Poissonian
dispersion of the \epeiso correlation 
is 
an important source of information. As discussed in Section 3 and shown
in Table 3, while the
correlation is very highly significant, the scatter of the data around the best
fit power--law exceed that expected by statistical fluctuations alone and
produces high values of \cqr. By fitting with a Gaussian the dispersion of the central 
values of log(\epi) around the best fit model, I obtain
$\sigma$$_{logE_{p,i}}$$\sim$0.2, while by fitting the whole data with the method
by \cite{Dagostini05}, which includes sample variance directly in the model, 
I obtain $\sigma$$_{logE_{p,i}}$ = 0.15$_{-0.4}^{+0.4}$ . A similar
scatter, even if computed with only the first of the two methods reported above,
is found for the \epi -- $L_{\rm iso}$ and \epi -- $L_{\rm peak,iso}$
correlations, see, e.g., \cite{Ghirlanda05a}. Intriguingly, the
\epega correlation shows instead a lower dispersion, of the order of
$\sigma$$_{logE_{p,i}}$$\sim$0.1 \citep{Ghirlanda04a,Nava06},
even if this correlation is based on a still low number of
events, it requires an estimate of $t_b$ in addition to \epo and $z$, 
it depends on jet model and circum--burst environment properties
(density, distribution) and there are possible outliers, as
discussed e.g. by \cite{Friedman05} and may be indicated by the lack of break in the 
X--ray afterglow light curve of some \swift GRBs with known redshift. 
A low 3--D dispersion characterizes
also the \epeisotb correlation, 
which is a kind of 
model--independent version of the \epega correlation \citep{Liang05,Nava06}.
The existence of both the 
\epeiso and \epega correlations 
is due to the fact that the collimation angles of GRBs are distributed over
a relatively narrow range of values; the lower dispersion of the
\epega correlation indicates that at least part of the 
scatter of the \epeiso 
correlation is due to the dispersion of jet opening angles. 
And indeed, the comparison
of the properties of the two correlations has been used, in addition to
the study of the relation between jet opening angle and radiated energy,
to infer the distribution of jet opening angles, as done, e.g., by 
\cite{Ghirlanda05c,Bosnjak06a,Donaghy05a}.
Very recently, it has been found evidence that a relevant contribution
to the dispersion of the correlation is due to temporal properties
of the prompt emission, like the half--width of the auto--correlation
function \citep{Borgonovo06} and the "high--signal" time scale
\citep{Firmani06}.
Other contribution to the scatter of the \epeiso correlation may come from
viewing angle effects, e.g. \cite{Levinson05}, the dispersion of the
parameters of the fireball and/or of the time scales 
(as discussed in Section 5.1 concerning 
synchrotron emission in internal shocks),
the inhomogeneous
structure of
the jet, e.g. \cite{Toma05}),
the possible presence of significant
amount of material in the circum--burst region, that would affect the estimates
of both \eiso and \epi with a global qualitative effect of steepening the 
correlation and increasing its dispersion \citep{Longo05}.
In general, several GRB population synthesis models, like those mentioned at
the beginning of this Section,
predict a scatter of the \epeiso correlation which depend
on the parameters values.

When investigating the above physical interpretations and implications, 
it is important to take into account that at least part of 
the dispersion of the \epeiso correlation could arise from
instrumental and other systematic effects in the
estimates of \epi and \eiso and their uncertainties.
As discussed e.g. by \cite{Lloyd00,Lloyd02}, data truncation effects, i.e. the
systematics introduced by the limited energy band of the detector, may 
affect significantly the estimate of \epo. Indeed, as discussed
in Section 2, typical GRB spectra
are characterized by a very smooth curvature and cover $\sim$3 
orders of magnitude or more in photon energy. Thus, unless the 
energy band extends from few keV to few MeV, only 
a portion of the spectrum can be detected by a single instrument,
which may cause a bias in the estimate of the spectral
parameters, especially when \epo is not far from the low or 
high thresholds. 
Also, both \sax ($\sim$2--700 keV) and HETE--2 (2--400 keV), 
in addition to be capable to
detect X--ray rich events and XRFs which could not be triggered by BATSE 
($\sim$25--2000 keV), showed
that X--ray emission below $\sim$30 keV of normal GRBs can last 
up to several tens of seconds more than in the hard X--ray energy band.
Thus, a GRB detector working at energies higher than few tens of keV, 
like BATSE, may have lost, for a fraction of GRBs, a substantial portion
of the soft X--ray emission, with a consequent overestimate of 
$\alpha$ and \epo .
These effects can indeed be seen when comparing the X-- and hard X--rays
duration and light curves of \sax \citep{Frontera00b,Amati02} and HETE--2 
\citep{Sakamoto05b} GRBs and the average spectral parameters estimated
by \sax/WFC+GRBM and BATSE \citep{Amati02,Jimenez01}
for those events revealed by both satellites. This is true, even if to a minor
extent, when comparing the best fit spectral models obtained
with HETE--2/FREGATE (7--400 keV) alone \citep{Barraud03}
with those obtained by jointly fitting HETE--2/WXM (2--30 keV) and
HETE--2/FREGATE data \citep{Sakamoto05b}.

Concerning \eiso, the main source of possible systematics is the extrapolation
to the 1--10000 keV cosmological rest--frame energy band of the spectral model 
obtained by fitting data in the instrument energy band (see Section 2).
Indeed, the (known) statistical uncertainties and (unknown) biases in 
the estimates of spectral parameters may affect significantly the estimate
of \eiso . This is particularly true for estimates based on spectra from
instruments with high energy bound at a few hundreds of keV, like HETE--2 
or \swift/BAT,
which in several cases cannot provide a reliable estimate of the 
high energy spectral index $\beta$ . In addition, the typical choice of 
computing \eiso in the 1--10000 keV rest--frame energy band may not be 
optimal for very soft events with values of \epi below a few tens of keV,
for which this method can likely lead to an underestimate of \eiso. Off
course, also the choice of the cosmological parameters for the computation 
of the luminosity distance, usually made by assuming values in the ranges given by the
so called "concordance cosmology" based on type Ia SNe and CMB 
measurements, affects the values of \eiso .

Finally, very recently \swift/XRT found evidence of X--ray flares following
the end of the prompt emission as detected by \swift/BAT, e.g.
\cite{Burrows05}. One possible
interpretation of these phenomena is that they are due to continued 
activity of the engine and/or late
internal shocks \citep{Fan05,Burrows05,Wu05,Perna06,Proga06} and thus can be considered 
part of the prompt emission. Given that
these events are typically soft and that their fluence can be a significant
fraction of that of the GRB, their non--detection (because of sensitivity)
by past and current GRB detectors may also have biased the
estimates of \eiso (under--estimate) and  of \epi (over--estimate).

\subsection{Short GRBs}

In the past, the \epeiso correlation has been studied
basing on data of long GRBs, given that no redshift information was available
for short GRBs. Nevertheless, thanks to the measurements performed by HETE--2
and \swift, in the last year it has been possible to detect afterglow
emission and possible optical counterparts and/or host galaxies for a few short
GRBs. As discussed in Section 3, I have included in the analysis reported in
this paper the two short GRBs with firm estimates of $z$ and \epo:
GRB050709 \citep{Villasenor05,Hjorth05} and GRB051221
\citep{Golenetskii05e,Berger05}. As can be clearly seen in Figure 3, these
events are outliers to the \epeiso correlation, even when taking into
account its extra--Poissonian dispersion. In addition, the spectral data of other
recently localized short GRBs with possible redshifts, 
GRB050509b \citep{Gehrels05,Pedersen05}, GRB050724 \citep{Krimm05,Berger05a}
and GRB050813 \citep{Sato05b,Berger05b},
indicate a likely inconsistence with the \epeiso
correlation, even though an estimate of \epo was not possible for these events.
These evidences confirm the expectations based on the
fact that short GRBs tend to form a separate class with respect to long GRBs
in the hardness--intensity plane, i.e. they tend to be weaker and harder,
and clearly shows the potentiality of the use of the \epeiso plane for
distinguishing different classes of GRBs and understanding their nature.
In particular, given that 
short and long GRBs partially overlap in the hardness--intensity and
hardness--duration planes, that a fraction of short GRBs show a softer extended
emission which can last tens of seconds, e.g. \cite{Norris06}, the possible
relevant impact of spectral/temporal trigger selection effects, it is sometimes 
difficult
to establish to which class a GRB belongs. Thus the \epeiso plane
may be a powerful tool under this respect, especially when combined e.g
with the lack of spectral lag and spectral evolution
which seems to characterize short GRBs
\citep{Norris06}.

While the commonly accepted hypothesis for the progenitors of long GRBs
is their association with the core collapse of massive fastly rotating 
stars (so called "hypernova" or "collapsar" models), based on their duration
(from few seconds up to hundreds of seconds), their huge \eiso (Figure 1),
their typical location inside blue galaxies with high star formation rate and the 
evidence of a metal--rich circum--burst environment (as inferred from 
absorption / emission features in prompt and afterglow emission
spectra), short GRBs  
are thought to originate from the coalescence of neutron star -- neutron star
or neutron star -- black hole binaries
(the "merger" scenarios). It has also been proposed that a fraction of them 
may be giant flare from extra--galactic
soft gamma repeaters, see, e.g., \cite{Meszaros02,Piran05} for reviews. The very
recent observations mentioned above, show that, as long GRBs, also
short GRBs lie at cosmological distances (0.1$<$z$<$1) but they are less
energetic (see Table 1 and Figures 1 and 3). Also, their host galaxies show
different morphologies and star forming activity: for instance, short GRB050724
(as possibly short GRB050509b) came from an
elliptical galaxy with low star formation, whereas short GRB050709  
was associated with an irregular late type star forming galaxy.
Both events were located towards the outskirts of their host galaxies.
All these properties match the predictions
of the merger scenarios (e.g. Belczynski et al. 2006 \nocite{Belczynski06}). 
Off course, these evidences concern still a very low number 
of short GRBs and thus it is premature to draw any definitive conclusion. Anyway,
the different nature of the progenitors between short and long GRBs
can help in understanding their different behavior in the \epeiso plane.
\cite{Ghirlanda04b} found, basing on BATSE data, that the emission properties
of short GRBs are similar to those of the first $\sim$1--2 s of long GRBs. This
could indicate that the central engine is the same for the two classes, but works
for a longer time in long GRBs. This would explain the low radiated energy
by short GRBs and, given that the emission would stop before the 
typical hard to soft evolution observed
for long GRBs, also their high \epi (with respect to their \eiso). The merger 
scenarios naturally explain the short life of the central engine, and thus the 
low radiated energy and the quitting of the emission before hard to soft 
spectral evolution.
In addition, they also predict a weak afterglow emission, as recently observed
\citep{Fox05}, because
of the cleaner and lower density circum--burst medium with respect to that
predicted by the hypernova scenarios for long GRBs, which would cause a very 
inefficient external shock. In the GRB scenarios
where most of the prompt emission of long GRBs is due to the external 
shock too, this naturally explains the lack of long lasting and softening
emission in short GRBs. Again, this would produce an high average \epi value
with respect to the radiate energy, and thus the
inconsistency with the \epeiso correlation holding for long GRBs. Finally, in order
to explain the very low or 0 spectral lag observed in short GRBs with
respect to long GRBs, \cite{Norris06} consider the hypothesis that the
typical $\Gamma$
of short events is several times that of long ones, and thus of the
order of $\sim$500--1000, as predicted e.g by the merger model of
\cite{Aloy05}. With such a high Lorentz factor, internal shocks are expected to
have a low efficiency, and indeed this is one possible 
explanation for the weakness and
softness of XRFs (e.g Barraud et al. 2005 \nocite{Barraud05}). 
If we assume the scenario proposed by e.g.
\cite{Ghirlanda03}, in which the spectrally hard emission 
observed in the first seconds of long GRBs
is due to (possibly Compton dragged) thermal emission from the photosphere and
the later emission to synchrotron processes occurring in internal shocks, 
the high $\Gamma$ would thus produce a short--hard emission 
possibly followed by a weak soft
component, as observed for several short GRBs \citep{Villasenor05,Norris06}.

\subsection{Sub--energetic GRBs and the GRB/SN connection}

As can be seen in Figure 3, in addition to the two short GRBs also
the prototype event for the GRB/SN connection,
GRB980425/SN1998bw, is characterized by values of \epi and \eiso completely
inconsistent with the \epeiso correlation holding for the other events. From an
observational point of view, this is a direct consequence of the fact that the event
is characterized by a fluence and a measured peak energy in the range of classical
long GRBs but, based on the commonly accepted association with SN1998bw,
it lies at a much lower distance
(z = 0.0085). 
Figure 3 shows that
also another event associated with a SN event, GRB031203, is characterized by
a value of \epi which, combined with its low value of \eiso,
makes it completely inconsistent with the correlation.
Given that GRB031203 is the most similar event
to GRB980425 under several points of view 
(although lying at a larger distance,
z $\sim$ 0.1), in particular the strong evidence of association with a SN event
(SN2003lw) and the low afterglow energy
inferred from radio observations \citep{Soderberg04,Sazonov04},
this inconsistency has been invoked as a further evidence of the existence
of a class of close sub--energetic GRBs.
However, the lower limit on \epi based on ISGRI data \citep{Sazonov04} 
and the \epi estimate based on Konus data (Table 2, Ulanov et al. 2005)
are currently debated based on the dust echo observed with XMM--Newton,
which could indicate a much softer prompt emission spectrum \citep{Watson06}. 
The sample of GRBs with most evidence of association with a SN include also
GRB030329 (SN2003dh) and GRB021211 (SN2002lt), which, in converse, are not 
sub--energetic and are
characterized by \epi and 
\eiso values fully consistent with the \epeiso correlation.
The fact the two closest and sub--energetic among those GRBs most 
clearly associated with a SN are 
outliers to the \epeiso correlation is intriguing. 
As in the case of short GRBs, these evidences show the potential use of
the \epeiso
plane
to distinguish among different sub--classes of GRBs, and have important implications for
GRB/XRF/SN unification models.
The most common explanations assume that
the peculiarity of these events is due to particular and uncommon 
viewing angles,
as proposed e.g. by \cite{Yamazaki03b} for GRB980425 and \cite{Ramirez05b} 
for GRB031203. Based on relativistic beaming and Doppler effects and
the assumption of a uniform jet, they find
that \epi $\propto$ $\delta$ and \eiso $\propto$ $\delta$$^{3}$, where $\delta$
is the relativistic Doppler factor (see Section 5.2). For large off--axis
viewing angles the different dependence of \epi and \eiso on $\delta$ would cause
significant deviations form the \epeiso correlation. 
An alternative explanation has been suggested by \cite{Dado05}
in the framework of the CB model \citep{Dar04}. In this scenario,
the \nufnu spectra of GRBs are characterized by two peaks, one at sub--MeV energies,
the normal peak following the \epeiso correlation, and one in the GeV--TeV range. 
When a GRB is seen very off--axis, the same relativistic
Doppler and beaming effects discussed above would shift the high energy peak 
at low energies, making it to be confused with the normal low energy GRB peak.
Very recently, off--axis scenarios have been seriously challanged by the detection
of a very close ($z = 0.0331$) and sub--energetic event, GRB060218/SN2006aj,
which, contrary to GRB980425, is characterized by \epi and \eiso values
fully consistent with the \epeiso correlation \citep{Amati06,Ghisellini06}. \\
Finally, \cite{Bosnjak06b}, based on BATSE data of GRBs with unknown redshift found
evidence that a part of GRB with indications of association with a SN 
are inconsistent with the
\epeiso correlation, as also the lag--luminosity relation \citep{Norris00},
for any value of $z$, 
which would confirm the existence of a peculiar class of
sub--energetic, SN--associated events. See, however, next Section for a discussion
of this method.

\section{The \epeiso correlation and GRBs with unknown redshift}

\subsection{Pseudo--redshifts, GRB luminosity function, cosmology}

The existence of correlations between intrinsic properties of GRBs,
emerged in the last years thanks to the increasing number of GRBs with known 
redshift, has
stimulated their use for the estimate of pseudo--redshifts for large 
samples of BATSE GRBs. In turn, the pseudo--redshifts estimates
have been used to compute the
luminosity of large samples of GRBs and infer their luminosity function.
This has been done mainly with the correlation between spectral lag and 
luminosity discovered by \cite{Norris00}, 
the variability -- peak luminosity correlation 
\citep{Reichart01,Reichart05,Guidorzi05}, 
and the \epi -- peak luminosity
correlation \citep{Yonetoku04}. In addition, the \epega and \epeisotb correlations,
given their low dispersion, have been used for the estimate of cosmological
parameters, e.g., \citet{Ghirlanda04c,Dai04,Liang05,Xu05}, in a way similar to
that used with type Ia SNe .
With respect to all the above correlations, the \epeiso correlation 
is based on a much larger sample of GRBs, as shown in this 
work, and is very highly significant. Also, differently from the 
\epega and \epeisotb correlation, which show a lower dispersion but 
an higher number of possible outliers (as discussed above),
the \epeiso correlation does not require the detection of a break in the afterglow 
light 
curve nor a modelization of the GRB jet and circum--burst environment 
(as needed for the \epega correlation). 
Thus, in principle the \epeiso can provide the most reliable
pseudo--redshift estimates.
The most straightforward method  
is to take the fluence
and spectral parameters of a GRB and compute \epi and \eiso, following the same
methods used by \cite{Amati02} or \cite{Ghirlanda04a}, for a grid
of $z$ values (say from 0.01 to 50). The pseudo--redshift range will then be 
given by the
values of z for which, accounting for the uncertainties on
\epi and \eiso and for the extra--Poissonian dispersion of the 
correlation (as quantified in Section 4 and discussed above), the
corresponding (\epi,\eiso) points are consistent with the \epeiso
correlation at a given level of significance. In practice, if $K$ and
$m$ are the normalization and index of the power--law best fitting
the \epeiso correlation ($\sim$100 and $\sim$0.5 if we assume the values 
determined in Section 4 by accounting for sample variance), \epi
and \eiso are the intrinsic peak energy and the isotropic radiated energy
at the redshift $z$, the significance of the deviation of the
(\epi,\eiso) point from the correlation is given by 
$ \Delta/\sqrt{\sigma_\Delta^2 + \sigma_{corr}^2}$, where
$\Delta = log(E_{p,i}) - log(K) - m \times log(E_{iso})$,
$\sigma_\Delta$ is the uncertainty on $\Delta$ computed 
from $\sigma_{logE_{p,i}}$ and $\sigma_{logE_{iso}}$ by error
propagation and $\sigma_{corr}$ is the extra--Poissonian dispersion
of the correlation (based on the sample variance analysis reported
in Section 4, one can for instance assume $\sigma_{corr}$=0.15).
Off course, to be more accurate one should also take into account the
uncertainties on $K$, $m$ and $\sigma_{corr}$; however, $K$ and $m$
are correlated and have low uncertainties (Table 3) and for 
$\sigma_{corr}$ one can conservatively assume the upper bound
of the 90\% c.l. interval (0.17).
This method, as can be seen by testing it on GRBs with known $z$ and
published spectral parameters, like e.g. the \sax events
included in the sample of \cite{Amati02} or the HETE--2 events analyzed
by \cite{Barraud03} and \cite{Sakamoto05b},
provides reliable but often large ranges of 
pseudo--z (or only lower limits). 
More precise $z$ estimates can be obtained with
redshift indicators partially based on the \epeiso correlation,
like the one developed by \cite{Atteia03}, which provides redshift
estimates accurate to a factor of $\sim$2 
and is currently used to estimate pseudo--redshifts of HETE--2 GRBs,
or its refined version proposed very recently by \cite{Pelangeon06}.
A caution on the use of these redshift estimators comes from the fact
that they are partially empirical and thus are not supported by a
complete understanding of the underlying physics.

\subsection{Tests and selection effects}

As already pointed out by \cite{Amati02}, given the relevance of the
\epeiso correlation, attention has to be paid to the possible
impact of selection effects.
Recently, two research groups \citep{Nakar05,Band05}, by analyzing
BATSE GRBs without
known redshift, inferred that $\sim$half \citep{Nakar05} or even $\sim$80\% 
\citep{Band05} of the whole
GRB population cannot satisfy the correlation for any values of redshift.
Thus, they conclude that strong selection effects are introduced in the various steps
leading from GRB detection to the final $z$ estimate 
and that we are measuring the redshift
of only those events that follow the correlation.
However, there are increasing evidences
that the possible impact of selection effects on the \epeiso correlation
and the possible number of outliers are much
lower than argued by these authors.
First, their conclusions have been questioned by several other 
authors \citep{Ghirlanda05c,Bosnjak06a,Pizzichini06}, 
who found instead that the \epo and
fluence values of most BATSE GRBs with unknown redshift are fully consistent
with the \epeiso correlation. The main source of 
discrepancy
between these two different conclusions lies in accounting or not for the observed
dispersion of the correlation and for the uncertainties in the \epo and fluence
values. When accounting for both in the way discussed in previous Section, it can
be found that only a very small fraction of BASTE GRBs with unknown redshift may be
considered outliers to the \epeiso correlation
\citep{Ghirlanda05c,Pizzichini06}. It has also to be noticed that
some of these authors assumed pseudo--redshifts derived by other correlations, like 
e.g. the lag--luminosity relation, which are necessarily not the $z$ values
corresponding to the (\epi,\eiso) with the minor deviation from the \epeiso relation
found with GRBs with known redshift. An important
effect that should be taken into account when using BATSE data to test the
correlation
is that, as discussed in previous Section, given to the
lack of coverage of the X--ray band below $\sim$25 keV, BATSE is likely to 
overestimate the \epi values at least for a fraction of GRBs. And indeed, 
\cite{Ghirlanda05c}, by fitting the (\epi,\eiso) points of 442 BATSE GRBs with
pseudo--redshifts derived by using the lag--luminosity correlation, find a
slope and dispersion consistent with the one obtained with GRBs with known redshift,
but a higher normalization, which could indicate a systematic overestimate of \epi
by BATSE with respect to \sax and HETE--2. 
I also note that the possible existence of an \epeiso correlation was suggested by
\citet{Lloyd00} based on BATSE data of GRBs without known redshift.

Secondly, as a check, I applied the pseudo--redshift estimate
method described above to a sample of 46 HETE--2 GRBs with spectral parameters
and fluences published by \cite{Barraud03} and \cite{Sakamoto05b}.
I considered only those events with both $\alpha$
and \epo constrained; I also discharged the few events in the sample
of \cite{Sakamoto05b} with $\beta$$>$$-$2. When the reported best fit
model is a cut--off power--law, I assumed for $\beta$ the typical value
of $-$2.5 . For those events contained in both samples, I took the 
spectral parameters from \cite{Sakamoto05b}, given that it reports
results based on data from both WXM and FREGATE, whereas the analysis 
of \cite{Barraud03} is based on FREGATE data only. The errors on
\epi and \eiso where derived basing on the errors on spectral parameters
and fluences reported in the two references. In the estimate
of \eiso as a function of $z$ I assumed a standard cosmology with
$\Omega_m$ = 0.3, $\Omega_\Lambda$=0.7 and $H_0$=65 km s$^{-1}$ Mpc$^{-1}$; 
for the \epeiso correlation
I assumed $K$=100, $m$=0.5 and $\sigma_{corr}$=0.15 . I find that
38 events ($\sim$83\%) are consistent within 1$\sigma$ with the correlation
(i.e. show $ \Delta/\sqrt{\sigma_\Delta^2 + \sigma_{corr}^2}$ $\le$1),
5 events ($\sim$11\%) are consistent within 2$\sigma$ and 2 ($\sim$4\%) within
3$\sigma$ (with values of 2.1 and 2.7). The only event with a substantial 
deviation from
the \epeiso correlation (3.82$\sigma$) is the short GRB020531.
It is also important to note that the pseudo--redshift ranges obtained
are fully consistent with the observed $z$ distribution.
These results show that at least most of HETE--2 GRBs with unknown $z$ are
potentially fully consistent with the \epeiso correlation holding
for long GRBs and give a further evidence that short GRBs do not follow
this correlation.
Thus, if there are selection effects in the sample of HETE--2 GRBs 
with known redshift, on which the \epeiso correlation is partly based, they 
are more likely due to detectors sensitivity as a function of energy
than to the subsequent processes
leading to redshift estimate. However, the fact that the distribution
of these GRBs in the fluence--\epo plane is consistent with that of BATSE 
events (Figure 16 of Sakamoto et al. 2005b) indicates that the possible
inconsistency of a 
fraction of BATSE GRBs with unknown redshift with the \epeiso
correlation may be due to an overestimate of \epo, as a consequence of
the effects discussed above and in Section 5.3 .\\
A third important issue concerns the fact that, as discussed in Section 4 and
shown in Figures 2 and 3, all \swift GRBs
with known redshift are consistent with the \epeiso correlation. This is
a strong evidence against the existence of relevant selection effects in the
updated sample of GRBs with known redshift on which the \epeiso correlation 
study here presented is based, because:
a) the burst detection sensitivity of \swift/BAT in $\sim$15--300 keV 
is comparable
to that of BATSE
and better than that of \sax and HETE--2 (see however Band 2003 \nocite{Band03} for
a comparison of the sensitivities of these different detectors as a function
of energy),
which in the past years contributed nearly all the measurements on which
the \epeiso correlation was based;
b) the very fast and precise localization capabilities of \swift/XRT
allowed to substantially reduce the selection effects also in the process
leading to redshift estimate (GRB precise localization, 
optical follow--up, optical afterglow and/or host galaxy
detection and spectroscopy). A drawback of BAT is that it can provide an estimate of
\epo only for a small fraction of events (15--20\% at most).\\
Basing on the above, it is unlikely that the \epeiso correlation is strongly
affected by selection effects. Anyway, the existence of sub--classes of GRBs not
following it cannot be excluded. As discussed in previous Section, the
analysis of the sample
of GRBs with known redshift and \epo shows that, in addition to short GRBs, a
class of sub--energetic events (like GRB980425 and possibly GRB031203)
with spectral--energy properties inconsistent with the correlation may exist. The
possibile existence of a fraction of long GRBs is also predicted 
by some GRB synthesis models, like, e.g., the one by \cite{Donaghy05a}
based on off--jet relativistic kinematics effects. 
Obviously, the most reliable test of the 
\epeiso
correlation and of the existence of one or more sub--classes of outliers
will come from the ongoing quick enlargement of the 
sample of GRBs with
known redshift and \epo both in number, thanks to \swift fast
and precise localizations, and in the coverage of the \epo--fluence
plane, as allowed by the GRB experiments with different energy bands and sensitivity 
presently operating, like \swift/BAT, HETE--2, Konus/Wind, Suzaku/WAM,
INTEGRAL/ISGRI, or that
will fly in the next future, like, e.g., those on board AGILE and GLAST.

\section*{Acknowledgments}

I am grateful to C. Guidorzi and R. Landi for their help in parts of the
analysis here reported.
I thank F. Frontera, C. Guidorzi and E. Montanari
for useful discussions
and their contribution to the development 
of the method used to fit the data
accounting for sample variance. 
I wish also
to thank  
G. Ghirlanda for useful discussion 
and the anonymous Referee
for his/her very careful reading of the paper and for comments and
suggestions
which improved its quality.

\label{lastpage}

\end{document}